\documentclass[usenatbib]{mn2e}

\usepackage{graphicx}
\usepackage{hyperref}

\newcommand\un[1]{{\,\rm #1}}
\newcommand\rs[1]{_\mathrm{#1}}
\newcommand\apj{ApJ}
\newcommand\aj{AJ}
\newcommand\aap{A\&A}

\title[Aspect angle in SN~1006]{Aspect angle for interstellar magnetic field in SN~1006}
\author[Petruk O., Dubner G., Castelletti G. et al.]{
O.~Petruk$^{1,2}$, G.~Dubner$^{3}$, G.~Castelletti$^{3}$, 
F.~Bocchino$^{4,5}$, D.~Iakubovskyi$^{6}$, \\ \\
{\LARGE \rm M.~G.~F.~Kirsch$^{7}$, M.~Miceli$^{4,5}$, S.~Orlando$^{4,5}$, I.~Telezhinsky$^{8}$}\\
$^{1}$Institute for Applied Problems in Mechanics and Mathematics, Naukova St.\ 3-b,
   79060 Lviv, Ukraine\\
$^{2}$Astronomical Observatory, National University, Kyryla and Methodia St.\ 8, 79008 Lviv, Ukraine \\
$^{3}$Instituto de Astronom\'{\i}a y F\'{\i}sica del Espacio (IAFE), CC 67, Suc. 28, 1428 Buenos Aires, Argentina\\
$^{4}$INAF — Osservatorio Astronomico di Palermo, Piazza del Parlamento 1, 90134 Palermo, Italy\\
$^{5}$Consorzio COMETA, Via S. Sofia 64, 95123 Catania, Italy\\
$^{6}$Bogolyubov Institute for Theoretical Physics, Metrologichna St. 14-b 03780 Kiev, Ukraine\\
$^{7}$European Space Agency (ESA), European Space Operations Centre (ESOC), Robert-Bosch-Str. 5, D-64293 Darmstadt, Germany \\
$^{8}$Astronomical Observatory, Kiev National Taras Shevchenko University, Observatorna St. 3, 04053 Kiev, Ukraine\\
}
\begin{document}

\date{Accepted .... Received ...; in original form ...}

\pagerange{\pageref{firstpage}--\pageref{lastpage}} \pubyear{2008}

\maketitle

\label{firstpage}

\begin{abstract}
  A number of important processes taking place around strong shocks in
   supernova remnants (SNRs) depend on the shock obliquity. The measured
   synchrotron flux is a function of the aspect angle between interstellar
   magnetic field (ISMF) and the line of sight. Thus a model of non-thermal
   emission from SNRs should account for the orientation of the ambient
   magnetic field.
   We develop a new method for the estimation of the aspect angle, based on
   the comparison between observed and synthesized radio maps of SNRs,
   making different assumptions about the dependence of electron injection
   efficiency on the shock obliquity.
   The method uses the azimuthal profile of radio surface
   brightness as a probe for orientation of ambient magnetic field 
   because 
   it is almost insensitive to the downstream distribution of magnetic 
   field and emitting electrons.
   We apply our method to a new radio image of SN~1006
    produced on the basis of archival VLA and Parkes data. The
    image recovers emission from all spatial structures with angular
    scales from few arcseconds to 15 arcmin.
   We explore different models of injection efficiency and
    find the following best-fitting values for the aspect angle of
    SN~1006: $\phi\rs{o}=70^\mathrm{o}\pm 4.2^\mathrm{o}$ if the injection is isotropic,
    $\phi\rs{o}=64^\mathrm{o}\pm 2.8^\mathrm{o}$ for quasi-perpendicular injection (SNR
    has an equatorial belt in both cases) and $\phi\rs{o}=11^\mathrm{o}\pm
    0.8^\mathrm{o}$ for quasi-parallel injection (polar-cap model of SNR). 
    In the last case, 
    SN~1006 is expected to have a centrally-peaked morphology  
    contrary to what is observed. 
    Therefore, our analysis provides some indication against the quasi-parallel injection model. 
\end{abstract}


\section{Introduction}

Non-thermal emission of supernova remnants (SNRs) is intensively
studied. It carries out important information about physics of strong
shocks, kinetics of cosmic rays and magnetic field properties. SNRs are
therefore observed with space and ground based observatories in
the X-ray, $\gamma$-ray and radio bands, and modeled with  {advanced codes for 
MHD and/or particle kinetics.}

The interstellar magnetic field (ISMF) creates different obliquity angles
with the shock normal in different places of the SNR surface. Efficiencies
of injection and acceleration, compression and/or amplification of ISMF
may depend on the shock obliquity. Therefore, in order to
model non-thermal emission, it is important to make assumptions about ISMF
orientation around these objects.

In many cases the ISMF, galactic in origin, may be assumed to be
rather uniform on the scales of SNR sizes. This is likely to be the case
of the bilateral supernova remnants (hereafter BSNRs,
\citet{kesteven-caswell-1987}; \citet{gaensler-1998}; 
but see \citet{Orletal07}
for implications of asymmetries in BSNRs), which are characterized by strong, 
opposite limbs. 
BSNRs with symmetric structure in radio images limit the orientation of 
the ISMF component in the plane of the sky. In the case of the BSNR archetype SN~1006, for instance,
the ISMF may be parallel to the symmetry axis, spanning from SE to NW,
or be perpendicular to it, running from NE to SW. The former corresponds
to barrel-like (equatorial belt) structure and the latter
to polar-cap structure \citep{Rotetal04}. However,
in order to draw reliable conclusions about the shock obliquity, it is {also
necessary} to consider the aspect angle -- the angle between ISMF and
the line of sight -- which is still unknown.

In the present paper, {we propose a new method for determination of
the aspect angle. We extract the azimuthal profiles of the synchrotron
surface brightness distribution in a given SNR and compare the observed
profiles with those synthesized from theoretical models, 
making different assumptions on the aspect angle and/or on the details of
 injection and acceleration of electrons. The ``true''
aspect angle is that of the best-fitting model. As a first application of
the method, we analyzed SN~1006.}

The plan of the paper is as follows. The method itself is explained
in Sect.~\ref{ISMF-method} where results of numerical simulations 
are presented. In Sect.~\ref{sec:radio} we use  
our method to determine the aspect angle toward SN~1006.
To achieve this, 
we produced a new radio image of the remnant, obtained from
re-analysis of archival VLA and Parkes data. An approximate formula 
for the azimuthal variation of the radio surface brightness is presented
in Sect.~\ref{ISMF-disc} where we also discuss effects of the shock modification 
on our results as well as some consequences on the injection model in 
SN~1006. 

\section{Aspect angle in bilateral SNRs: the method}
\label{ISMF-method}

\begin{figure*}
 \centering
 \includegraphics[width=17truecm]{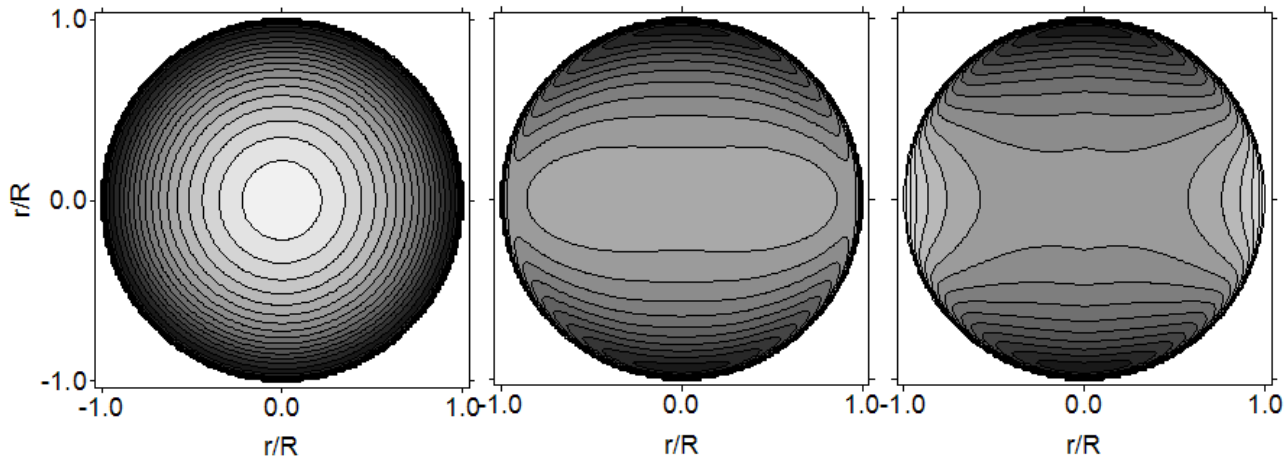} 
 \caption{Surface brightness distributions (in linear scale) 
 of the radio emission of adiabatic SNR 
 for different aspect angles: $\phi\rs{o}=0\degr$ ({\bf a}), 
 $60\degr$ ({\bf b}), $90\degr$ ({\bf c}). 
 Uniform ISM, uniform ISMF and isotropic injection are assumed. 
 In cases ({\bf b}) and ({\bf c}), the component of ISMF in the plane 
 of the sky is parallel to the horizontal axis. 
 In all three plots, the levels of brightness are spaced in the same way. 
                }
 \label{fig-a}
\end{figure*}

The method to determine the aspect angle 
$\phi\rs{o}$ between ISMF and the line of sight on the basis of radio emission in SNRs, 
can be deduced from the simulated radio maps produced
for a Sedov SNR expanding through a uniform ISM and ISMF, presented
in Fig.~\ref{fig-a}. As shown in the figure, the {\it azimuthal}
profile of the surface brightness is sensitive to the aspect angle:
it is constant for $\phi\rs{o}=0\degr$ (Fig.~\ref{fig-a}a: ISMF is
directed towards the observer) and it is steepest for $\phi\rs{o}=90\degr$
(Fig.~\ref{fig-a}c: ISMF is in the plane of the sky). Thus, comparison
of an observed azimuthal profile with theoretical ones allows
one to conclude about the aspect angle.

The radio surface brightness $S$ at some `point' of SNR image is

\begin{equation}
 S\propto \int K B^{(s+1)/2}dl
 \label{SN1006mf-eq1}
\end{equation}

\noindent
where $s$ and $K$ are the index and the normalization of the electron
energy spectrum, respectively, $B$ is the magnetic field (MF) strength
and the integration is along the line of sight within the volume of
SNR. For a Sedov SNR in a uniform ISM and uniform ISMF, the downstream
distributions of $K$ and $B$ are self-similar, i.e. may be written in the form 
$K=K\rs{s}(\Theta\rs{o})\bar K(\bar r)$,
$B=B\rs{s}(\Theta\rs{o})\bar B(\bar r,\Theta\rs{o})$ where 
$r$ is the radial distance from the centre of SNR, 
$\Theta\rs{o}$ is the shock obliquity angle between ISMF and the shock normal,
index 's' marks the immediate post-shock
values and the upper bar marks the downstream variables normalized
at their own values at the shock front \citep{Rey98,Pet06}.

The post-shock magnetic field is generally a subject of compression, 
and under conditions of efficient cosmic ray acceleration, amplification.  
The magnetic field can be expressed as  
$B\rs{s}=A\rs{B}(\Theta\rs{o})B\rs{o}$ where $A\rs{B}$ is a product of the compression 
factor $\sigma\rs{B}$ and an amplification factor. 
Since it is unknown whether magnetic field amplification depends on the 
shock obliquity, we assume it to
be independent of $\Theta\rs{o}$. In this case, the variation of the
post-shock MF with obliquity is only determined by the compression \citep{Rey98},
$B\rs{s}\propto\sigma\rs{B}(\Theta\rs{o})B\rs{o}$, where 

\begin{equation}
 \sigma\rs{B}(\Theta\rs{o})=\left(\frac{1+\sigma^2
 \tan^2\Theta\rs{o}}{1+\tan^2\Theta\rs{o}}\right)^{1/2},
 \label{sigmaB}
\end{equation}

\noindent
$\sigma=4$ is the shock compression ratio {for unmodified shocks (changes 
in this prescription and our results due to the shock modification 
are discussed in Sect.~\ref{ISMF-disc})}.

At the shock, the normalization, $K\rs{s}\propto \varsigma(\Theta\rs{o})$ where
$\varsigma$ is the injection efficiency defined as the
fraction of accelerated electrons. 
There are three alternatives for dependence of injection
efficiency $\varsigma$ on obliquity of the shock typically
considered in the literature: isotropic injection (i.e. $\varsigma$
independent of $\Theta\rs{o}$), quasi-parallel ($\varsigma\propto
\cos^2\Theta\rs{s}$) or quasi-perpendicular ($\varsigma\propto
\sin^2\Theta\rs{s}$) injection 
\citep{reyn-fulbr-90}. Therefore, the injection efficiency
either decreases (quasi-parallel) or increases (quasi-perpendicular) with 
increasing obliquity, or it is independent of $\Theta\rs{o}$. The MF compression
factor increases with $\Theta\rs{o}$.

In symmetric, bilateral SNRs (like SN~1006\footnote{Note
that SN~1006 is symmetric with respect to the axis between the two radio
lobes. However, the lobes appear slanted and converging to the SE
(see \citet{Orletal07}, for a possible explanation of this kind of
asymmetry).}), the possible orientations of the ISMF \emph{in the plane of
the sky} are limited. Namely, ISMF may be parallel to the symmetry axis
or perpendicular to it. For example, in SN~1006, if the injection is
isotropic or quasi-perpendicular, the bright limbs correspond to
the magnetic ``equator''  {(equatorial belt)} and ISMF should be oriented in SE-NW direction.
If injection prefers quasi-parallel shocks, {the bright limbs of}
SN~1006 are two polar caps and MF should be oriented in the NE-SW direction. 
In other words, the
model of injection determines the orientation of the plane-of-the-sky
component of ISMF.

There are theoretical expectations that injection is higher at parallel shocks 
\citep{Ellison-et-al-1995,Volk-Ber-injobliq-2003}. 
In contrast, observational evidence seems to argue for isotropic or 
quasi-perpendicular injection in 
SNRs \citep{reyn-fulbr-90,Orletal07}. We consider all
three possibilities in this work.


We synthesized a number of radio
surface brightness maps of a Sedov SNR evolving in a uniform ISM and
uniform ISMF. We used the model of \citet{Rey98} with some
extentions presented in \citet{Pet06}. This model is able to account 
for the evolution of $K\rs{s}\propto V^{-b}$ and $A\rs{B}\propto
V^{\beta/2}$ where $b$ and $\beta$ are constant and $V$ 
is the shock velocity. These relations reflect an eventual evolution of 
the injection efficiency and MF amplification at the shock. 

Radio maps of the model SNR were calculated for a range of indices: 
$s$ from $2$ to $2.2$, $b$ from $-3/2$ to $2$, $\beta$ from $0$ to $2$. 
For each set of parameters, we produced 
a series of images by changing the aspect angle $\phi\rs{o}$ 
from $0$ to $\pi/2$. For the image with 
$\phi\rs{o}=\pi/2$ in each series, we found the radius 
of projection $\varrho\rs{Smax}$ which corresponds to the position of the maximum
brightness $S\rs{max}=S(\varrho\rs{Smax},\varphi\rs{Smax})$ (by definition
$\varphi\rs{Smax}=\pi/2$ for this position)\footnote{The 
radius $\varrho\rs{Smax}$ 
is almost the same for any aspect angle, except for $\phi\rs{o}<30^\mathrm{o}$ 
in the quasiparallel model.
In the case of quasiparallel injection, the angle 
$\phi\rs{o}=30^\mathrm{o}$ 
roughly separates cases with a centrally peaked morphology 
($\phi\rs{o}<30^\mathrm{o}$) and a bilateral one ($\phi\rs{o}>30^\mathrm{o}$).}. 
We then traced the surface
brightness $S(\varrho\rs{Smax},\varphi)$ at this $\varrho\rs{Smax}$ for azimuthal
angles $\varphi$ from 0 to $\pi/2$ and plotted these distributions normalized by
the brightness $S(\varrho\rs{Smax},\varphi\rs{Smax})$.  

\begin{figure*}
\centering
\includegraphics[width=17.7truecm]{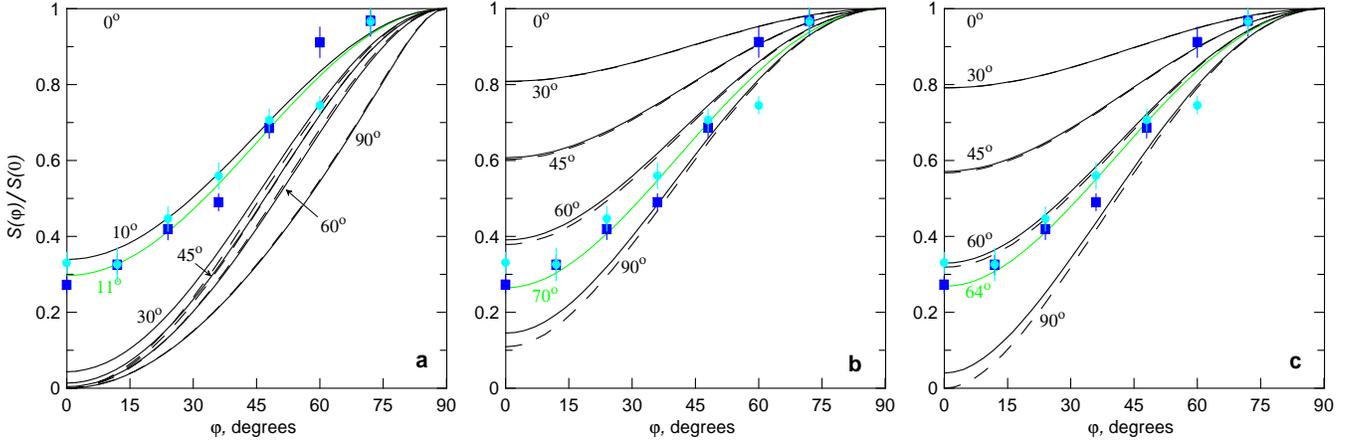}
\caption{Azimuthal variation of the radio surface brightness
$S(\varrho\rs{Smax},\varphi)$ for different aspect angles $\phi\rs{o}$.
Numerical results (solid lines) are normalized to
$S(\varrho\rs{Smax},\pi/2)$. Calculations are done for $b=0$,
$\beta=0$, $s=2.2$; the plots however are almost the same for different
values of $b$, $\beta$ and $s$. The models of injection are quasi-parallel  
({\bf a}), isotropic ({\bf b}) and quasi-perpendicular ({\bf c}).
$\bar \varrho\rs{Smax}=0.98$ ({\bf a}) and  
$\bar \varrho\rs{Smax}=0.97$ ({\bf b}, {\bf c}). 
Experimental data for SN~1006 are shown for region I (in blue) 
and II (in cyan). They are
measured at $1.5\un{GHz}$ between 12 and 14 arcmin from the centre 
and are normalized to the maximum
value of 1-$\sigma$ errors. 
Azimuthal profiles given by the approximate formula (\ref{ISMF:azimuthal}) 
are shown by dashed lines.} 
\label{fig-s}
\end{figure*}

Fig.~\ref{fig-s} shows the plots for different aspect angles, assuming $b=0$, 
$\beta=0$, $s=2.2$, and considering the three models for obliquity dependence of
injection efficiency. 
We found a reasonable result, namely that all {the
azimuthal profiles of surface brightness} which we obtained are almost
insensitive (within $10-20\%$) to the values of $s$, $b$ and $\beta$
(at least in the case of uniform ISM and uniform ISMF assumed
here). That means the azimuthal profiles of radio
brightness are almost independent of the shapes of the distributions of
relativistic electrons and magnetic field downstream of the shock. Such
stability allows one to safely use the method proposed here for
determination of the aspect angles from radio maps of SNRs if
the ISM and ISMF in which SNR expands can be considered to be mostly
uniform.


\section{The aspect angle in SN~1006}
\label{sec:radio}

The method is applicable to SNRs expanding in almost uniform ISM and
ISMF. BSNRs, particularly SN~1006, are ideal candidate targets of this study. 
In this SNR, uniform conditions are likely to be achievable since SN~1006 is located over 
500 pc above the galactic plane. In addition, the SE edge of the SNR exhibits a near-spherical shape, 
a good argument for expansion of the shock into a uniform ISM.

\subsection{Radio data}

We now apply our method for the determination of the aspect angle to
SN~1006 considerly the best example of a symmetric, bilateral SNR. To this end
we produced a new radio image of SN~1006 at $\lambda\sim 20$ cm.

This image was
produced on the basis of archival VLA\footnote {The Very Large Array of
the National Radio Astronomy Observatory is a facility of the National
Science Foundation operated under cooperative agreement by Associated
Universities, Inc.} data obtained in October 1991, February 1992 and
July 1992 in the hybrid AnB, BnC and CnD configurations, respectively. The
observations in the AnB configuration were carried out at 1370 and 1376
MHz, while the observations in the BnC and CnD arrays were performed at 1370
and 1665 MHz. The data corresponding to the more compact configurations
of the VLA, BnC and CnD, were published as a part of an expansion study
of SN~1006 \citep{Mofetal93}, but not the data from 
the AnB configuration, 
which provides the highest angular resolution of southern sources. The
new interferometric image is  produced on the basis of 4 hours per
configuration (the maximum possible taking into account the elevation
restrictions for this source when observed from the northern hemisphere)
and recovers emission from all spatial structures with angular scales
between a few arcsec and 15 arcmin.

All data were processed using the MIRIAD software
package. To avoid the diffraction effects produced
by point sources present in the field, for each of the brightest
sources we imaged a small region around, and the clean components were
Fourier transformed and subtracted from the visibilities. The residual
visibilities, containing all the source structure except for the offending
point sources were then imaged and the point sources were added back
into the SN~1006 image in the image plane. To recover flux density
contribution from structures on angular scales larger than 15 arcmin
(which is important in this case since SN~1006 is $\sim$30 arcmin in
diameter) we added single dish observations acquired in 2002 with the
Parkes 64 m radiotelescope placed in Australia. Also, since the primary
beam of the VLA (the half-power beamwidth of a single VLA antenna)
at $\lambda$20 cm is 32 arcmin, comparable to the size of the source,
a correction was applied to the interferometric data taking into account
the attenuation introduced near the primary beam edge.

\begin{figure}
 \centering
 \includegraphics[width=8.8truecm]{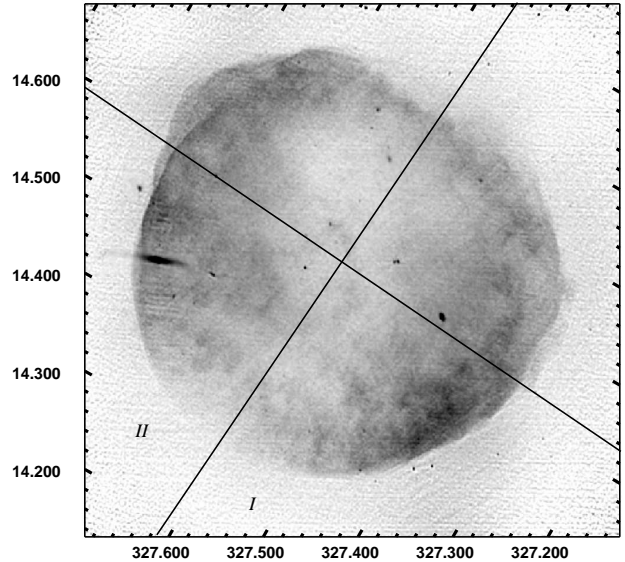}
 \caption{Radio image of SN~1006, in linear scale (galactic coordinates). 
 Axes outline regions I and II used in our analysis.
                }
 \label{interfSD}
\end{figure}

The final image has a synthesized beam of $7^{\prime\prime}.7 \times
4^{\prime\prime}.8$, position angle 8$^\circ$.3,  and an rms noise
of $1 \times 10^{-4}$ Jy/beam. When combined with Parkes single-dish
data, the total recovered flux is S=14.9 Jy, in excellent agreement
with previous estimates from Green's (2006) catalogue 
of SNRs\footnote{\url{http://mrao.cam.ac.uk/snrs/}}. The new image is
presented in Fig.~\ref{interfSD}. The accuracy of the field point sources
serves for comparison  to appreciate the quality of the new image that
resolves the SNR radio features down to the same fine spatial 
resolution\footnote{After submission of our paper, a new radio image
of SN 1006 has been reported by 
\citet{Cassam2008SN1006}. This image, based on 
new observations, has an angular resolution of 
$\sim$ 6\arcsec $\times$ 
9\arcsec, comparable to that achieved in this work. It has a better
noise level (20 $\mu$Jy/beam); however, it lacks the large spatial 
scale structures that were recovered in our image with the addition
of single dish data.}.

\subsection{Results}
 \label{SN1006mf:res}

To determine the MF orientation we only considered the SE half of SN1006 
(regions I and II on Fig.~\ref{interfSD}) because this part is quite spherical 
and therefore is more appropriate for comparison with the numerical results 
obtained in Sect.~\ref{ISMF-method} for a SNR in uniform ISM.
From the radio map of SN~1006, we extracted the radial brightness profiles 
(along radii of the SNR projection separated on $\Delta \varphi =12$ degrees). 
The experimental radial distributions are subject to pixel-to-pixel variations. 
In order to lower the possibility of error due to fluctuations in observational data, 
we calculated the averages of brightness and 1-$\sigma$ errors within 
12 to 14 arcmin from the centre of SN~1006 (where the maximum in radial 
distribution of the surface brightness is located). 

Experimental data are compared with the theoretical results on
Fig.~\ref{fig-s}. The estimated aspect angle $\phi\rs{o}$ differs
much for the polar-cap and the equatorial-rim models of SN~1006.
From the numerical simulations,
the best-fitted aspect angle is $\phi\rs{o}=70^\mathrm{o}\pm
4.2^\mathrm{o}$ for isotropic injection, $\phi\rs{o}=64^\mathrm{o}\pm 2.8^\mathrm{o}$ for
quasi-perpendicular injection and $\phi\rs{o}=11^\mathrm{o}\pm 0.8^\mathrm{o}$ for
quasi-parallel injection. 
Considering an isotropic injection and equatorial-rim model for SN~1006, 
\citet{Reyn96} found a similar aspect angle, $\phi\rs{o}=60^\mathrm{o}$.

\section{Discussion}
\label{ISMF-disc}

\subsection{Basic processes determining the azimuthal brightness profiles}

We found that the azimuthal profiles of the radio surface brightness of the 
Sedov SNR are almost independent of the shapes of the distributions of
relativistic electrons and magnetic field downstream of the shock. 
Close to the shock, the azimuthal variations of the radio brightness 
of the SNR may therefore be approximately described as (Appendix A): 
\begin{equation}
 S\rs{\varrho}(\varphi)\propto 
 \varsigma\big(\Theta\rs{o,eff}(\varphi,\phi\rs{o})\big)\ 
 A\rs{B}\big(\Theta\rs{o,eff}(\varphi,\phi\rs{o})\big)^{(s+1)/2}
 \label{ISMF:azimuthal}
\end{equation}
where
\begin{equation}
 \cos\Theta\rs{o,eff}=\cos\varphi\sin\phi\rs{o}
 \label{ISMF:angles:isotr}
\end{equation}
for isotropic or quasi-perpendicular injection and 
\begin{equation}
 \cos\Theta\rs{o,eff}=\sin\varphi\sin\phi\rs{o}
 \label{ISMF:angles:paral}
\end{equation}
for quasi-parallel injections. (Eq.~(\ref{ISMF:azimuthal}) is not valid for 
a centrally-brightened SNR.)


In order to compare the approximate azimuthal profiles 
given by this formula with the numerical results, 
we plotted them on Fig.~\ref{fig-s} with dashed lines. 
This figure shows that (\ref{ISMF:azimuthal}) can be 
used as an approximation for 
the azimuthal distributions of the radio surface brightness in SNRs 
evolving in uniform medium and uniform magnetic field.

An important consequence of (\ref{ISMF:azimuthal}) is that 
there are {\it two basic processes} which determine the {\it azimuthal}
distribution of the radio surface brightness in SNR evolving in
uniform ISM and uniform ISMF: electron injection and compression (and/or
amplification) of ISMF on the shock.

\subsection{Modified shock}

The estimations in Sect.~\ref{SN1006mf:res} were obtained for a shock compression 
ratio 
$\sigma=4$. Modified shocks may be responsible for larger compression,
$\sigma=7$. Modeling the situation out of the Bohm limit, we may 
use such compression factor together with (\ref{sigmaB}).
Essentially, the estimated aspect angle does not change
in this case. 
Namely, it becomes $\phi\rs{o}=65\degr\pm
4.2^\mathrm{o}$, $64^\mathrm{o}\pm 2.8^\mathrm{o}$ and 
$11^\mathrm{o}\pm 0.8^\mathrm{o}$ for isotropic,
quasi-perpendicular and quasi-parallel injection respectively. 
Thus, even if the shock is modified but not in the Bohm limit, 
the estimation for the aspect angle
obtained under the assumption of an unmodified shock, may well be valid.

Our model is based on a classical MHD approach where the parallel and perpendicular 
components of the ambient magnetic field are compressed in different ways, as it is given 
by (\ref{sigmaB}) 
(e.g. \citet{Korob1960} or 
\citet[Chapter 7.4]{Korob1991} for the self-similar solution of the problem of 
the strong point explosion in a 
constant magnetic field). This approach might not be valid in a limit in which the
turbulent magnetic field dominates, as in case of the very efficient non-linear particle 
acceleration consistent with Bohm diffusion. The quasi-parallel theory assumes in this 
case that the turbulence is produced ahead of the shock, not downstream. The compression 
of the (already turbulent) magnetic field then does not depend on the original obliquity 
\citep{Volk-Ber-injobliq-2003,Ber-Ksenof-SN1006-2002}. \citet{Rakowski-2008} argue 
that shocks of different initial obliquity 
subject to magnetic field amplification become perpendicular immediately 
upstream. 
We may model such situation 
assuming $\sigma\rs{B}(\theta\rs{o})$ is a constant instead of using (\ref{sigmaB}) and 
$\bar B(\bar r,\Theta\rs{o})=\bar B(\bar r,\pi/2)$. 

The estimations of the aspect angle in this case are 
$58^\mathrm{o}\pm 2.8^\mathrm{o}$ and 
$11^\mathrm{o}\pm 0.9^\mathrm{o}$ 
for quasi-perpendicular and quasi-parallel injection respectively.  
Isotropic injection produces constant azimuthal profiles in this case, 
i.e. the modeled SNR looks like a rim for any aspect angle. 

The estimated aspect angles in the assumption of the Bohm limit are close to those 
obtained for an unmodified shock, except for the isotropic injection. 
Isotropic injection, 
together with the assumption that the downstream magnetic field is independent of obliquity 
is not able to reproduce bilateral morphology of SNR. There is no azimuthal variation of 
the radio surface brightness in this case, in agreement with (\ref{ISMF:azimuthal}). 

\subsection{Injection efficiency and obliquity}

It is worth emphasizing that our analysis may have some
implications for the model of injection efficiency. 

\begin{figure}
 \centering
 \includegraphics[width=6.3truecm]{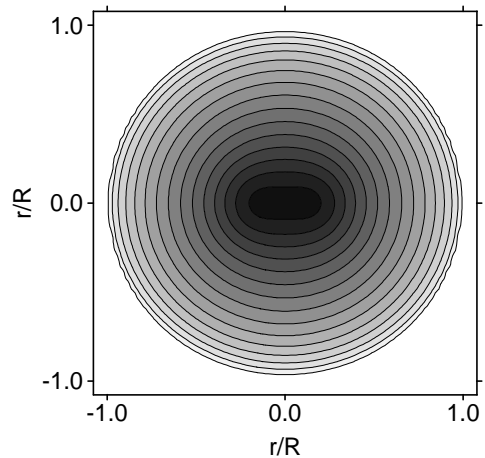}
 \caption{Surface brightness distribution (in linear scale) of SNR 
 with the same parameters as on Fig.~\ref{fig-s}. 
 The surface brightness distribution is calculated
 assuming the quasi-parallel model of injection and the 
 aspect angle $\phi\rs{o}=11\degr$. 
 The component of ISMF in the plane 
 of the sky is parallel to the horizontal axis. 
 The levels of brightness are spaced in the same way as on the 
 Fig~\ref{fig-a}. 
                }
 \label{fig-mapa}
\end{figure}

Our argument against the quasi-parallel injection is the morphology 
of SN~1006 it should have in uniform ISM and uniform ISMF. 
Fig.~\ref{fig-mapa} shows an image of SN~1006 in case of
the quasi-parallel injection ($\varsigma\propto \cos^2\Theta\rs{s}$) and
aspect angle $\phi\rs{o} = 11^\mathrm{o}$. Since the ambient magnetic
field should be almost aligned with the line of sight and injections
prefer parallel shocks (`polar caps' directed toward and away from
observer), the brightness distribution of SN~1006 should be centrally
brightened (with one or two radio `eyes' within thermal X-ray rim), 
contrary to what is observed.

Our argument seems 
therefore to disagree with polar-cap morphology, and favour a
NW-SE orientation of interstellar magnetic field around this SNR. 

\citet{Rotetal04} suggested a criterion, which is considered as 
an argument on behalf of the polar-cap model for SN~1006. 
Namely, if SN~1006 has an equatorial rim the observer should see some emission 
between the bright limbs. Numerically, the value of the parameter 
$R\rs{\pi/3}$ defined as a ratio between total power coming from the interior 
and that from the limbs should be $R\rs{\pi/3}>0.5$ if SN~1006 is a barrel. 
The most likely explanation for the smaller value of $R\rs{\pi/3}$ is that the 
visible limbs are polar caps \citep{Rotetal04}. 
The value of this ratio is $R\rs{\pi/3}\approx 0.7$ in radio 
\citep{Rotetal04}. Therefore, based on the $R\rs{\pi/3}$ criterion alone, the 
radio data itself cannot give preference neither to equatorial-rim nor to 
polar-cap model of SN~1006. However, $R\rs{\pi/3}\leq 0.3$ in X-rays 
\citep{Rotetal04}, prefering therefore the latter 
model. 

To this end, our argument against polar caps are in contradiction 
with $R\rs{\pi/3}$-criterion applied to X-ray data. 
Note, that the $R\rs{\pi/3}$-criterion is obtained for cylindrical source of 
isotropic 
emission. It would be interesting to see how deviations from this assumption 
may affect the criterion. 
Our models assume a uniform ISM and uniform ISMF. 
Could it be possible to reproduce 
the bilateral morphology (i.e. to obtain $\phi\rs{o}>30^\mathrm{o}$) 
with quasiparallel injection if one considers a gradient in the ISMF? 
Assuming the contrast of the ISMF between the NE and SE regions can be 
determined by the relationship 
$B\rs{SE}/B\rs{NE}\simeq \left(S\rs{SE}/S\rs{NE}\right)^{2/3}$, 
a ratio of $4$ for $\phi\rs{o}=45^\mathrm{o}$, and  
$20$ for $60^\mathrm{o}$, 
could make the azimuthal profile of radio emission comparable to the observed one. 
Further investigation, including multi-dimensional modeling of SN~1006, 
as recent VHE $\gamma$-ray data 
could help us to understand the nature of morphology of this SNR. 

\section{Conclusions}
\label{ISMF-concl}

Radio images of SNRs can be a useful probe to determine the orientation 
of ISMF around SNRs. We develop a
method for determination of the aspect angle between ISMF and the line
of sight, through the comparison between radio observations and model
calculations. It is based on a property of azimuthal variation of radio
surface brightness. Namely, the azimuthal profile is almost insensitive to
the downstream evolution of magnetic field and relativistic
electrons. In contrast, it is sensitive to the obliquity dependences of
both the injection efficiency and the compression (amplification) factor
of ISMF. The obliquity angle, the azimuthal angle and the aspect angle
are geometrically related. The simple expression (\ref{ISMF:azimuthal})
together with certain assumptions about the model of injection (where isotropic
or quasi-perpendicular injection appears to be suggested by a number of 
arguments 
\citep{reyn-fulbr-90}) may be used in order to
fit the observed azimuthal profile and to approximately estimate the
aspect angle.

We applied our
method to determine the aspect angle in SN~1006. A new radio image of
SN~1006 obtained from archival VLA and Parkes data is reported. The image
has very good angular resolution and has contributions from all spatial scales recovered. 
The use of the numerical modelling together with the new observations reported here allows us to find
the best-fitting value of the aspect angle in case of unmodified shock: 
$\phi\rs{o}=70^\mathrm{o}\pm 4.2^\mathrm{o}$
for the case of isotropic injection, $\phi\rs{o}=64^\mathrm{o}\pm 2.8^\mathrm{o}$
for quasi-perpendicular injection (equatorial-belt model of SNR in both
cases) and $\phi\rs{o}=11^\mathrm{o}\pm 0.8^\mathrm{o}$ for quasi-parallel injection
(polar-cap model). 
The angles estimated under assumption of modified shock are quite close 
to those given above. 

There are some limitations from our results on the model of injection. 
In quasi-parallel and 
quasi-perpendicular models, aspect angles are expected (and found) 
to be similar in both modified and unmodified shocks 
because the obliquity variation of injection dominates the obliquity 
variation of ISMF compression/amplification. 
Rather flat observed azimuthal radio profile strongly prefers $\phi\rs{o}<30^\mathrm{o}$ 
if injection is assumed quasi-parallel (Fig.~\ref{fig-s}a), i.e. for polar-cap model of SN~1006. 
In this case, SN~1006 should be centrally brightened, 
contrary to what is observed. Next, if one assumes the shock in SN1006 is so strongly 
modified that compression of ISMF is independent of obliquity (Bohm limit), 
then 
the only possible injection model is quasi-perpendicular, because isotropic injection 
and constant $A\rs{B}$ results in constant azimuthal profiles of the radio surface 
brightness, again contrary to what is observed. 

Rejection of the quasi-parallel injection model in SN~1006 means that 
the initial ISMF is directed from SE to NW and SN~1006 has a barrel-shaped, 
rather than polar-cap, morphology. 

With our results, we come to a puzzling issue which should be investigated 
in the future: 
\citet{Rotetal04} $R\rs{\pi/3}$-criterion applied to the X-ray 
emission seems 
to exclude the equatorial model of SN~1006 while the analysis of the azimuthal 
radio profiles seems to be against of the polar-cap scenario.

\section*{Acknowledgments}

The draft of the paper was improved thanks to critical remarks of 
anonymous referee.
FB, MM, and SO acknowledge Consorzio COMETA under the PI2S2 Project, a
project co-funded by the Italian Ministry of University and Research
(MIUR) within the Piano Operativo Nazionale `'Ricerca Scientifica,
Sviluppo Tecnologico, Alta Formazione' (PON 2000-2006). OP and DI acknowledge 
the program 'Kosmomikrofizyka' of National Academy of Sciences (Ukraine). 
GD and GC are members of CIC-CONICET (Argentina) and acknowledge support 
of PIP-CONICET 6433. 
DI acknowledges the support from the INTAS project No. 05-1000008-7865. 
IT acknowledges the support from the INTAS YSF grant No. 06-1000014-6348.


\appendix
\section[]{Approximate formula for azimuthal variation of the radio surface brightness in Sedov SNR}

Here we derive an approximate formula for 
azimuthal variation of the radio surface brightness. 
This formula allows one to avoid detailed numerical simulations and may 
be useful in situations where the approximate estimation of the aspect angle 
is reasonable. 
In addition, the formula allows us to have deeper insight in the main factors
determining the azimuthal variation of the radio surface brightness in SNRs. 

The downstream distributions of $K$ and $B$ 
in a Sedov SNR in uniform ISM and uniform ISMF are 
\begin{equation}
 K\propto\varsigma(\Theta\rs{o})\bar K(\bar r),\quad 
 B\propto A\rs{B}(\Theta\rs{o})\bar B(\bar r,\Theta\rs{o}). 
\end{equation}
If one neglects the small
differences in downstream distributions of the parallel and perpendicular
components of $B$ \citep[Fig.~1]{Rey98}, then 
\begin{equation}
 \bar B(\bar r,\Theta\rs{o})\approx \bar B(\bar r).
\end{equation}

The obliquity angle $\Theta\rs{o}$ is different for each radial sector of 3-D object. 
It is determined, for any position within SNR, by the set 
$(\varphi,\bar r/\bar \varrho,\phi\rs{o})$. 
Integration along the line of sight gathers information from different radial 
sectors, with different obliquities. 
Let us determine the `effective' obliquity angle by the relation 

\begin{equation}
 \Theta\rs{o,eff}(\varphi,\phi\rs{o})=\Theta\rs{o}(\varphi,1,\phi\rs{o}). 
\end{equation}

\noindent
Actually, $\Theta\rs{o,eff}$ 
for a given azimuth equals to the obliquity angle for a sector with the same azimuth 
lying in the plane of the sky (i.e. in the plane being perpendicular to the line of sight 
and containing the center of SNR). 
$\Theta\rs{o}$ varies around $\Theta\rs{o,eff}$ during integration along the line of sight. 
The closer $\varrho$ to the edge of SNR projection the smaller the range for 
variation of $\Theta\rs{o}$ and more accurate is our approximation. 
(Actually, we used $\varrho$ 
corresponding to maximum in radial brightness distribution which happens rather 
close to the shock.) 

Let us consider the azimuthal profile of the radio brightness $S\rs{\varrho}$ 
at a given radius $\varrho$ from the centre of the SNR projection. 

With the use of $\Theta\rs{o,eff}$, 
the azimuthal variation of the radio brightness for fixed $\varrho$ may 
approximately be written from (\ref{SN1006mf-eq1}) as

\begin{equation}
 S\rs{\varrho}\propto \varsigma(\Theta\rs{o,eff}) A\rs{B}(\Theta\rs{o,eff})^{(s+1)/2} 
 \!\!\int^{1}_{\bar \varrho}\! {\bar K(\bar r) \bar B(\bar r)^{(s+1)/2}\bar r d\bar r\over 
 \sqrt{\bar r^2-\bar \varrho^2}}.
 \label{ISMF:eq2}
\end{equation}

\noindent
The integral in (\ref{ISMF:eq2}) is the same for any azimuthal
angle $\varphi$. The azimuthal variation of the radio brightness is therefore
approximately determined by

\begin{equation}
 S\rs{\varrho}(\varphi)\propto 
 \varsigma\big(\Theta\rs{o,eff}(\varphi,\phi\rs{o})\big)\ 
 A\rs{B}\big(\Theta\rs{o,eff}(\varphi,\phi\rs{o})\big)^{(s+1)/2}.
 \label{ISMF:azimuthal:appendix}
\end{equation}

\noindent
The relation between the azimuthal
angle $\varphi$, the obliquity angle $\Theta\rs{o,eff}$ and the aspect angle
$\phi\rs{o}$ is as simple as

\begin{equation}
 \cos\Theta\rs{o,eff}=\cos\varphi\sin\phi\rs{o}
\end{equation}

\noindent
for assumption of isotropic or quasi-perpendicular injection 

\begin{equation}
 \cos\Theta\rs{o,eff}=\sin\varphi\sin\phi\rs{o}
\end{equation}

\noindent
for quasi-parallel injections 
(the relations are different because we define the azimuth angle 
$\varphi\rs{Smax}=\pi/2$ at the position of the maximum brightness). 

The accuracy of the approximation (\ref{ISMF:azimuthal:appendix}) 
is shown on Fig.~\ref{fig-s}. 
The azimuthal profiles is sensitive to $\varrho$ 
in quasi-parallel case for aspect angles less than about $30^\mathrm{o}$, 
i.e. for SNR with centrally-brightened radio morphology 
($\varsigma\propto \cos^2\Theta\rs{o}$ 
and, for small aspect angles, 
$\Theta\rs{o}\rightarrow\pi/2$ on the perifery of SNR and thus
$\varsigma\rightarrow 0$ there). Therefore, 
the formula (\ref{ISMF:azimuthal:appendix}) does not give correct profiles 
in the case of quasi-parallel injection for $\phi\rs{o}<30^\mathrm{o}$, 
unless $\bar\varrho\rightarrow 1$. 


\label{lastpage}

\end{document}